\documentclass[twocolumn,superscriptaddress,amssymb,amsmath,aps,longbibliography,nofootinbib,prl,floatfix]{revtex4-2}
\usepackage{graphicx}
\usepackage{dcolumn}
\usepackage{bm}
\usepackage{float}
\usepackage{amsmath}
\usepackage{amssymb}
\usepackage{color}
\usepackage{subeqnarray}
\usepackage[utf8]{inputenc}
\usepackage{textgreek}
\usepackage[english]{babel}
\usepackage{mathrsfs}
\usepackage[bb=boondox]{mathalfa}
\usepackage{xcolor} 
\usepackage{gensymb}
\usepackage{amsbsy}
\usepackage{latexsym,epsfig,graphicx}
\usepackage{subfigure}
\usepackage{amsfonts}
\usepackage{amsmath}
\usepackage{xspace}
\usepackage{epstopdf}
\usepackage{comment}
\usepackage{array,booktabs}
\usepackage{babel}
\usepackage{multirow}
\usepackage[colorlinks=true, pdfstartview=FitV, linkcolor=red, citecolor=blue, urlcolor=blue]{hyperref}
\usepackage[normalem]{ulem}

\usepackage{cleveref,braket,xcolor}

\pdfoutput=1

\begin{document}

\title{AI Scientist Discovered Robust Quantum Control}

\author{
Jiahao~Yan,$^{1,*}$
Xuanying~Lai,$^{1,*}$
Xiaoshui~Lin,$^{1,*}$
Yizhou~Wang$^{1}$,
Changyu~Yao$^{1}$,
Guanghui~He$^{1}$,
Thomas~Lewer$^{1}$,
Chong~Zu$^{1,2,\dag}$
Chuanwei~Zhang,$^{1,2,\dag}$
\\
\medskip
\normalsize{$^{1}$Department of Physics, Washington University, St. Louis, MO 63130, USA}\\
\normalsize{$^{2}$Center for Quantum Leaps, Washington University, St. Louis, MO 63130, USA}\\
\normalsize{$^*$These authors contribute equally to this work}\\
\normalsize{$^\dag$To whom correspondence should be addressed; E-mail: zu@wustl.edu, chuanwei.zhang@wustl.edu}\\
}

\date{\today}

\begin{abstract}
Protecting quantum coherence is a central theme of modern quantum technologies. 
Carefully designed control pulses can extend coherence by suppressing disorder and unwanted interactions. 
However, designing robust pulse sequences remains challenging in interacting multi-level systems, where multiple independent error channels must be controlled in the presence of experimental imperfections.
Here we introduce QORA, a multi-agent artificial intelligence framework that integrates theoretical reasoning, automated experimental calibration and control for closed-loop discovery of quantum control sequences.
Applied to a dense ensemble of spin-1 nitrogen-vacancy centers in diamond, QORA discovers AID-ROD, a series of AI-discovered sequences that nearly double the measured spin coherence time relative to the state-of-the-art human-designed sequence DROID-C3PO.
Subsequence analysis by human researchers reveals a hierarchy of error suppression: 18-pulse blocks suppress linear and quadratic onsite disorder and common-mode drive-amplitude errors, while their arrangement over a longer cycle reduces dipolar interactions, providing a physical explanation for the observed improvement and robustness in AID-ROD.
These results demonstrate AI-guided experimental discovery of robust quantum control and identify a design strategy for protecting coherence in interacting multi-level systems beyond qubits.
\end{abstract}

\maketitle
\raggedbottom

\section{Introduction}\label{sec1}

Preserving quantum coherence is essential for quantum technologies \cite{Suter2016CoherenceProtection, Onizhuk2025SpinDecoherence}.
It has enabled reliable quantum computation \cite{Jain2024Penning,Bland2025Coherence,Manetsch2025Coherence}, storage \cite{Wang2019QuantumMemory,Knaut2024Memory,Liu2026Memory} and sensing \cite{Panda2024Coherence,Yang2025Cat,Lu2026Freezing}, as well as controlled study of many-body dynamics  \cite{Bluvstein2021ManyBody,Andersen2025Thermalization,Manovitz2025Coarsening}.
Dynamical decoupling (DD) addresses this challenge through periodical control sequences that average out the unwanted evolution \cite{Hahn1950SpinEcho,Carr1954CP,Meiboom1958CPMG,
Viola1999DD,Ryan2010Robust,Souza2011RobustDD}.
While the pulses suppress disorder and interaction, they also introduce errors \cite{Lang2019Detuning,Farfurnik2015OptimizingDD,Souza2011RobustDD,Ishikawa2018FiniteWidth,Lancaster2026PulseErrors}.
Their accumulation makes coherence protection a balance between averaging unwanted dynamics and maintaining accurate control~\cite{Choi2020Robust,Zhou2024Qudit}.

This balance becomes particularly demanding in multi-level quantum systems, where additional degrees of freedom introduce independent error channels and more complex interaction dynamics that must be addressed simultaneously.
For instance, in spin-1 nitrogen-vacancy (NV) center ensembles, linear and quadratic disorder proportional to $S_i^z$ and $(S_i^z)^2$, coexist with dipolar interactions and pulse imperfections.
Average Hamiltonian theory (AHT) \cite{Haeberlen1968AHT} and symmetry-based design \cite{Choi2020Robust} provide systematic rules for robust pulse design.
DROID-C3PO, for example, uses a hierarchical construction to protect spin-1 coherence while suppressing disorder and control errors during finite pulses \cite{Zhou2024Qudit}.
Yet these rules do not uniquely determine the pulse ordering and timing that best preserve coherence.
Sequences satisfying the same leading cancellation conditions can accumulate distinct higher-order residue errors, whose effect depends on the competing strengths of disorder, interactions, and control imperfections~\cite{Zhou2023HigherOrder}.
Finding effective sequences therefore requires exploring a large space of pulse orderings and timings, using analytical constraints to guide the search and experimental feedback to identify designs that remain robust under realistic conditions.

Conventional machine-learning algorithms broaden the search for pulse sequences, but their effectiveness depends strongly on the choice of loss function and the accuracy and computational cost of the underlying simulations~\cite{Quiroz2013GeneticDD, Peng2022DeepRL, Tong2025EmpiricalDD, Zhang2026NearLimit}.
Recent advances in Large language model (LLM)-based agents enable direct interaction with experimental control systems, allowing agents to coordinate workflows and operate scientific instruments with human oversight~\cite{Boiko2023Coscientist, Bran2024ChemCrow, Chen2026XrayScientist, Cao2025QuantumAgents, Vriza2026Instruments,Reed2026Principles}.
These agents can also synthesize scientific knowledge and formulate experimentally testable hypotheses~\cite{Swanson2025VirtualLab, Lu2026AIScientist, Gottweis2026CoScientist, Ghareeb2026Robin, Miao2026Paper2Agent}.
Together, these capabilities raise a fundamental question: can AI agents discover new physical mechanisms and principles through direct interaction with experimental systems?
Quantum control provides a concrete setting for exploring this question, connecting AI-driven hypothesis generation and autonomous experimentation to the physical interpretation of newly discovered control protocols.

\begin{figure*}[t]
    \centering
    \includegraphics[width=\textwidth]{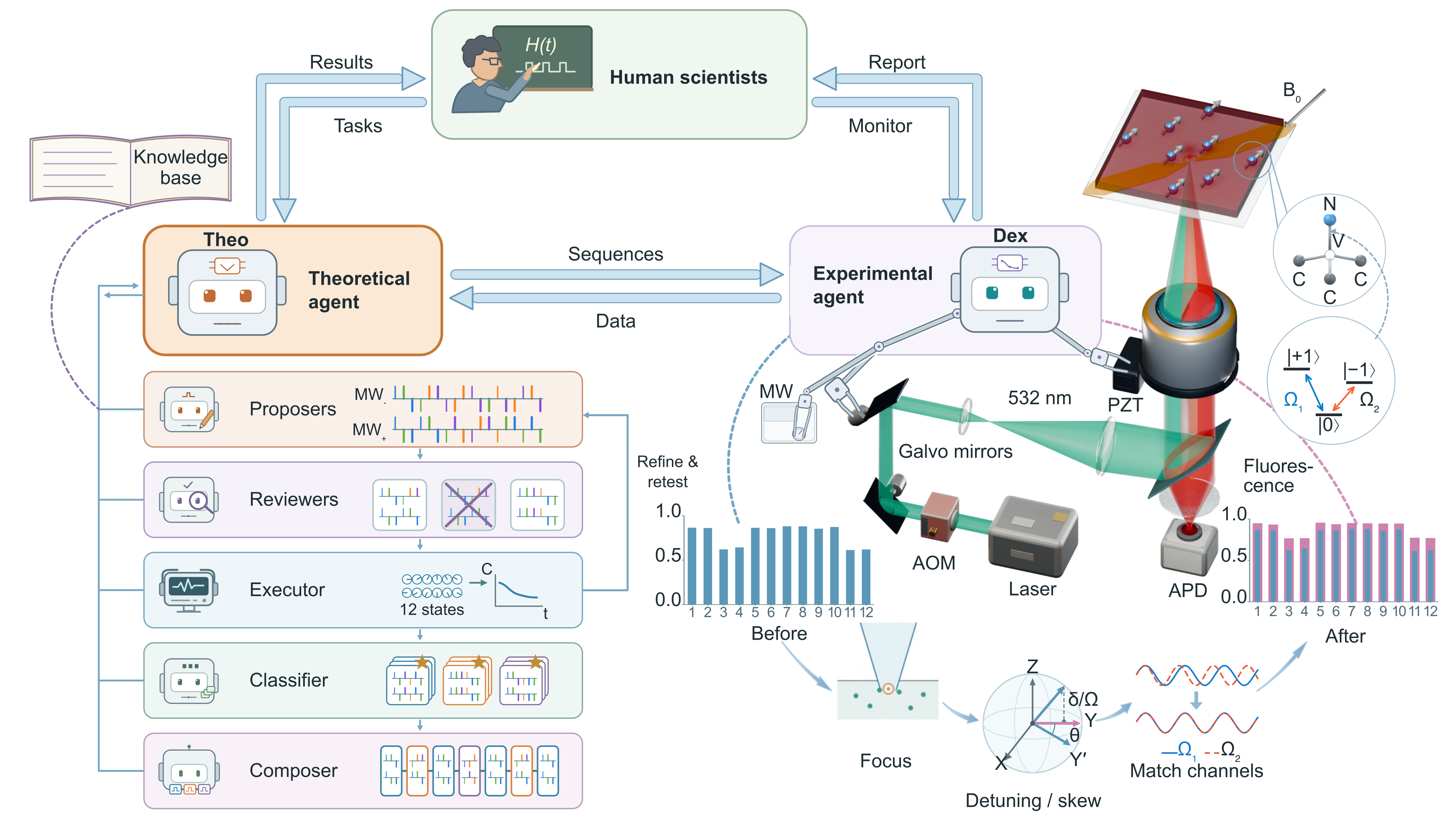}
    \caption{Closed-loop sequence design and calibration with QORA. Human scientists set scientific goals, assign tasks to the theoretical agent Theo and monitor the experimental agent Dex as needed.A knowledge base supports Theo’s scientific workflow. Proposers define families of 18-pulse blocks; reviewers assess structural redundancy and proposed comparisons; and the executor computes coherence contrasts for twelve initial states. The classifier groups qualifying blocks by pulse structure, and the composer assembles seven blocks derived from a selected parent block into a complete cycle. Numerical results guide proposal refinement and retesting. Then Theo sends sequences to Dex, which implements them on the NV-center apparatus and returns experimental data. The apparatus combines optical excitation, fluorescence readout and microwave control. Dex calibrates focus, detuning, the X/Y drive skew angle and microwave-channel matching as needed before repeating measurements. The bar charts compare estimated fidelities for 12 states, which are $|\psi_{mn,\phi}\rangle=(|m\rangle+e^{i\phi}|n\rangle)/\sqrt{2}$,
where $(m,n)\in\{(0,+1),(0,-1),(+1,-1)\}$ and
$\phi\in\{0,\pi/2,\pi,3\pi/2\}$, before and after calibration. Experimental data guide further sequence design, closing the theory–experiment loop.}\label{<fig1>}
\end{figure*}

Here, we introduce the Quantum Optimization and Research Agent (QORA), a multi-agent artificial intelligence system which links sequence design to automated experiments on NV ensembles (Fig.~\ref{<fig1>}).
QORA automates optical and microwave control, experimental execution and data analysis, while monitoring experimental drift and triggering recalibration as needed.
Under human-defined goals and supervision, QORA discovers AID-RODs (AI-Discovered Robust decoupling sequences), a family of robust dynamical decoupling sequences for an interacting spin-1 NV ensemble.
Its representative member, AID-ROD-1, nearly doubles the measured coherence time relative to the state-of-the-art human-designed protocol DROID-C3PO, with advantages persisting across a wide range of controlled parameters (Fig.~\ref{<fig2>}).
Subsequent analysis by human researchers reveals a hierarchy of error suppression, in which 18-pulse blocks suppress both linear and quadratic on-site disorder and common-mode drive-amplitude errors, while longer cycles reduce dipolar interactions (Fig.~\ref{<fig3>}).
These results demonstrate how collaboration between AI and human researchers can generate new scientific knowledge by discovering improved quantum control protocols and revealing the physical mechanisms underlying their performance.

\section{QORA: Quantum Optimization and Research Agent}\label{sec2}

Powered by GPT-6-Astra and Sorol, QORA connects human scientists, a theoretical agent, Theo, and an experimental agent, Dex, in a closed feedback loop for quantum control discovery (Fig.~\ref{<fig1>}). Human scientists define the scientific goals and assign tasks to Theo, while reviewing experimental reports from Dex as needed.
Theo coordinates five specialized sub-agents that propose, review, simulate, classify and assemble control sequences. Drawing on a knowledge base compiled from the literature on dynamical decoupling and Hamiltonian engineering, Theo refines candidate sequences in response to simulation results. Dex uses the GPT API to interact with instrument-control software, implementing the selected sequences, acquiring measurements and calibrating the apparatus as needed. The resulting data return to Theo to guide further sequence design.

For spin-1 search, Theo organizes sequence design into block discovery and composition
(Fig.~\ref{<fig1>}), drawing on the hierarchical construction from literature~\cite{Zhou2024Qudit}. 
The design strategy reflects the separation of energy scales in our NV ensemble, where microwave driving error and local disorder are much stronger than dipolar couplings. 
Theo therefore searches for 18-pulse blocks that suppress these noise channels, then combines transformed versions of a selected block into a longer cycle to reduce dipolar interactions. 
A complete cycle contains seven 18-pulse blocks, giving 126 pulses in total. 
Each pulse comprises two simultaneous microwave tones addressing the two transitions between the three spin-1 sublevels, ($|0\rangle\leftrightarrow|+1\rangle$) and ($|0\rangle\leftrightarrow|-1\rangle$), with each tone’s phase independently selected from a ($\pi$/4) grid. 
Cyclic shifts and tone-specific phase offsets provide the degrees of freedom for block composition. 

Proposers design internal pulse substructures and choose their arrangement and relative phases within a block. 
Each proposal defines a sequence family and specifies which features are held fixed and which are varied to test a hypothesis for suppressing disorder. 
Reviewers assess structural redundancy and whether
the proposed comparisons test the hypothesis before admitting proposals for numerical evaluation.
The executor evaluates how effectively each candidate sequence preserves
the coherence of twelve initial states,
$|\psi_{mn,\phi}\rangle=(|m\rangle+e^{i\phi}|n\rangle)/\sqrt{2}$,
where $(m,n)\in\{(0,+1),(0,-1),(+1,-1)\}$ and
$\phi\in\{0,\pi/2,\pi,3\pi/2\}$. 
Their evolution is computed using exact diagonalization.

\begin{figure*}[htbp!]
    \centering
    \includegraphics[width=\textwidth]{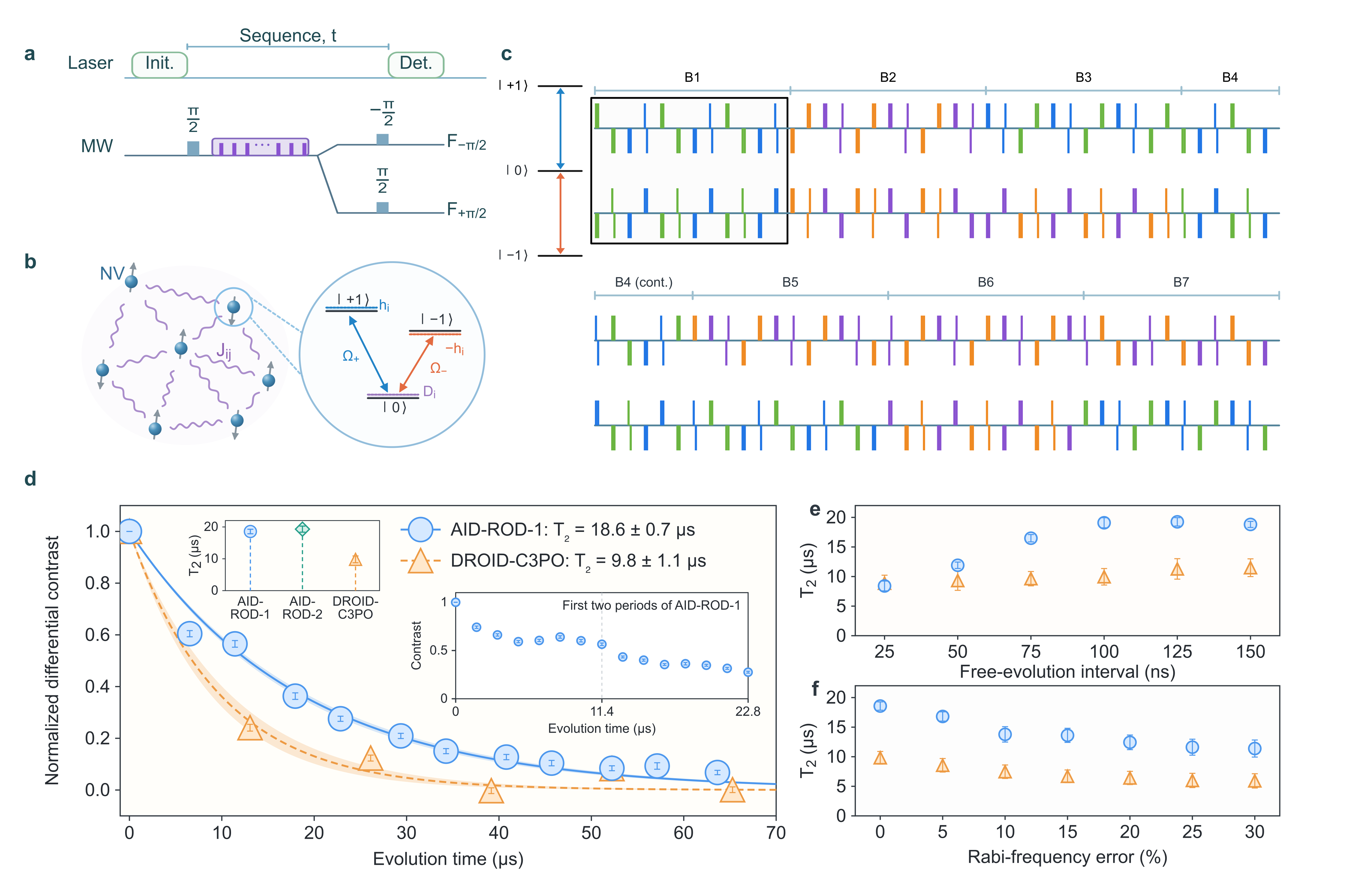}
    \caption{ \textbf{a}, Protocol for measuring spin coherence. A 532-nm laser pulse initializes the spin in $|0\rangle$, and a $\pi/2$ pulse prepares the state to be preserved. After dynamical decoupling, a $\pm\pi/2$ analysis pulse maps the remaining coherence onto the $|0\rangle$ population. Optical readout gives the differential contrast $C(t)=\left(F_{-\pi/2}(t) - F_{+\pi/2}(t)\right)/\left(F_{-\pi/2}(0) - F_{+\pi/2}(0)\right)$, where $F_{\pm\pi/2}(t)$ denotes the fluorescence measured after the corresponding analysis pulse following a control sequence of duration $t$.
\textbf{b}, Dephasing mechanisms targeted by the sequence: NV--NV dipolar interactions, local detuning disorder $h_i$ and zero-field splitting disorder $D_i$. The detuning shifts the two transitions with opposite signs. Simultaneous driving with equal Rabi frequencies, $\Omega_{+} = \Omega_{-}$, enables control of the spin-1 system.
\textbf{c}, AID-ROD-1, proposed by QORA to suppress these dephasing mechanisms. One cycle comprises seven 18-pulse blocks (B1--B7), displayed across two rows. Each block contains three identical copies of a six-pulse unit. The upper and lower tracks address the $|0\rangle \leftrightarrow |+1\rangle$ and $|0\rangle \leftrightarrow |-1\rangle$ transitions, respectively. Orange, green, purple and blue denote phases of $0^\circ$, $45^\circ$, $90^\circ$ and $135^\circ$. Narrow and wide bars represent $\pi/2$ and $\pi$ pulses; bars below the baseline carry an additional phase of $180^\circ$. Here, $\tau = 100\,\mathrm{ns}$ and the cycle duration is $T = 11.4\,\mu\mathrm{s}$.
\textbf{d}, AID-ROD-1 exhibits slower coherence decay than DROID-C3PO , indicating a longer coherence time. The contrasts measured after each block for the first two cycles of AID-ROD-1 are presented in the right inset. The upper-left inset compares $T_2$ for AID-ROD-1 and AID-ROD-2, the two sequences proposed by QORA, with that of DROID-C3PO.
\textbf{e}, Dependence of $T_2$ on the free-evolution interval, probing the balance between faster decoupling and accumulated pulse errors.
\textbf{f}, Robustness to drive-amplitude errors, tested by varying the Rabi frequency. AID-ROD-1 retains more coherence than DROID-C3PO  under these errors.}\label{<fig2>}.
    
\end{figure*}

At a pulse spacing of 100~ns, a block qualifies for classification only if its ensemble-averaged readout differential contrast $C(t)$ across the twelve states have a mean above 0.8 and a minimum above 0.7, where $C(t)$ is measured as in Fig.~\ref{<fig2>}a.
State-resolved responses and their corresponding pulse parameters are returned to the proposers to guide subsequent revisions.

The classifier groups qualifying blocks by pulse structure and selects representative parent blocks for composition.
For each parent block ($B$), the composer constructs a seven-block cycle by jointly selecting the cyclic starting point of each block and a uniform phase offset for each of its two microwave tones.
To assess the coherence improvement of composition, the resulting state-resolved decay curves are compared with those obtained by repeating the parent block seven times ($B^7$).
Complete sequences are also benchmarked against DROID-C3PO under matched simulation conditions. 

Selected sequences are passed to Dex for experimental evaluation (Fig.~\ref{<fig1>}), where Dex converts them into control waveforms for an arbitrary waveform generator and measures the resulting coherence decay profile.
To ensure reproducible measurements, Dex autonomously calibrates the optical alignment, microwave driving frequencies, amplitudes and  phases, while monitoring 12 benchmark spin states to detect and compensate experimental drifts. 
The measured coherence curves are returned to Theo to guide subsequent sequence design, closing the autonomous discovery cycle.

\section{Coherence improvement in spin-1 NV ensemble}\label{sec3}

Our experiments are performed on a DNV-B14 bulk diamond containing NV centers with density 1.2~ppm. Each NV center hosts a spin-1 ground state, with three sublevels, $\lvert 0\rangle$ and $\lvert \pm 1\rangle$, forming a natural qutrit Hilbert space relevant to higher-dimensional quantum information processing, control and sensing~\cite{Fu2022TwoQutrit,zu2014experimental, Nagata2018Holonomic, Zhou2024Qudit,Barry2020Sensitivity}. 
The $\lvert \pm 1\rangle$ states are separated from $\lvert 0\rangle$ by a zero-field splitting of $\Delta \approx 2\pi \times2.87\,\mathrm{GHz}$, while an external magnetic field of $B_0 = 123\,\mathrm{G}$ aligned with one of the four NV crystallographic orientations lifts the $\lvert \pm 1\rangle$ degeneracy by $2\pi\times0.687\,\mathrm{GHz}$, allowing selective addressing of the aligned NV subensemble. 
Green laser excitation optically polarizes the spins into $\lvert 0\rangle$, followed by a $\pi/2$ microwave pulse that prepares the desired coherent initial state. The ensemble then evolves under a selected pulse sequence with programmable amplitude, phase, frequency, and pulse spacing (Fig.~\ref{<fig2>}a).
The remaining coherence is measured through differential fluorescence readout using complementary $\pi/2$ and $-\pi/2$ pulses.

The coherence of the NV ensembles is limited by local disorder and dipolar interaction (Fig.~\ref{<fig2>}b).
In the rotating frame and employing the secular approximation, the effective spin-1 Hamiltonian is given by
\begin{equation}
\begin{aligned}
H &= \sum_i \left[ h_i S_i^z + D_i (S_i^z)^2 \right]\\
&\quad+ \sum_{ij} J_{ij} \left[
S_i^z S_j^z
- \frac{1}{2} H_{ij}^{XY,0+}
- \frac{1}{2} H_{ij}^{XY,0-}
\right],
\label{eq:NV_Hamiltonian}
\end{aligned}
\end{equation}
where $h_i$ and $D_i$ characterize linear and quadratic on-site disorder, arising primarily from local magnetic-field variations and lattice strain respectively~\cite{Zhang2026dephasingmechanism}.
The dipolar interaction, with pairwise coupling coefficients $J_{ij}$, comprises longitudinal Ising terms $S_i^zS_j^z$ and flip-flop terms defined by $H_{ij}^{XY,0\pm} \equiv \left[\bigl(\lvert\pm1\rangle\langle0\rvert\bigr)_i \bigl(\lvert0\rangle\langle\pm1\rvert\bigr)_j +\mathrm{H.c.}\right]$, which couple $\lvert0\rangle_i\lvert\pm1\rangle_j$ and $\lvert\pm1\rangle_i\lvert0\rangle_j$, describing excitation exchange between the two NV spins.
In experiments, local disorder and dipolar interactions continue
to act during finite-duration microwave pulses. The spins
therefore evolve under $H_{\rm total} = H+H_c(t)$, where
\begin{equation}
\begin{aligned}
H_c(t)=\frac{1}{2}\sum_i\Big[
&\Omega_+(t)e^{i\varphi_+(t)}
\bigl(\lvert+1\rangle\langle0\rvert\bigr)_i\\
&+\Omega_-(t)e^{i\varphi_-(t)}
\bigl(\lvert-1\rangle\langle0\rvert\bigr)_i
+\mathrm{H.c.}\Big].
\end{aligned}
\label{eq:NV_control}
\end{equation}
Here $\Omega_\pm(t)$ and $\varphi_\pm(t)$ denote the angular
Rabi frequencies and phases of the drives addressing
$\lvert0\rangle\leftrightarrow\lvert\pm1\rangle$, respectively. Robust sequence design must therefore account for the effects of
disorder and dipolar interactions during the control pulses.

QORA discovers AID-RODs, a family of dynamical decoupling sequences with cycle durations comparable to DROID-C3PO. 
Each cycle contains seven 18-pulse blocks, and each block consists of three identical copies of a six-pulse unit, giving 126 pulses per cycle. 
These design features improve coherence preservation: AID-ROD-1 achieves $T_2=18.6\pm0.7~\mu\mathrm{s}$, compared with $9.8\pm1.1~\mu\mathrm{s}$ for DROID-C3PO (Fig.~\ref{<fig2>}d). Each block also implements an approximate identity propagator, allowing coherence to be sampled at block boundaries rather than only after a complete cycle. This finer sampling is illustrated in the right inset of Figure~\ref{<fig2>}d. 
A second experimentally tested sequence, AID-ROD-2, achieves a comparable coherence time, $T_2=19.3\pm1.0~\mu\mathrm{s}$ (upper-left inset of Fig.~\ref{<fig2>}d).

We next access the robustness of the coherence enhancement to variations in the free-evolution interval between pulses and to the pulse errors.
The dependence on the pulse spacing reflects a trade-off between faster dynamical decoupling and accumulated pulse errors. 
Across the measured intervals, AID-ROD-1 consistently outperforms DROID-C3PO (Fig.~\ref{<fig2>}e).
We further vary the Rabi-frequency error from $0\%$ to $30\%$ via adjusting microwave drive amplitude. 
Although coherence times decrease with increasing error for both protocols, AID-ROD-1 retains its advantage throughout the tested range (Fig.~\ref{<fig2>}f).

\section{Physical interpretation of AI-designed sequences}\label{sec4}

\begin{figure*}[t]
    \centering
    \includegraphics[width=1\linewidth]{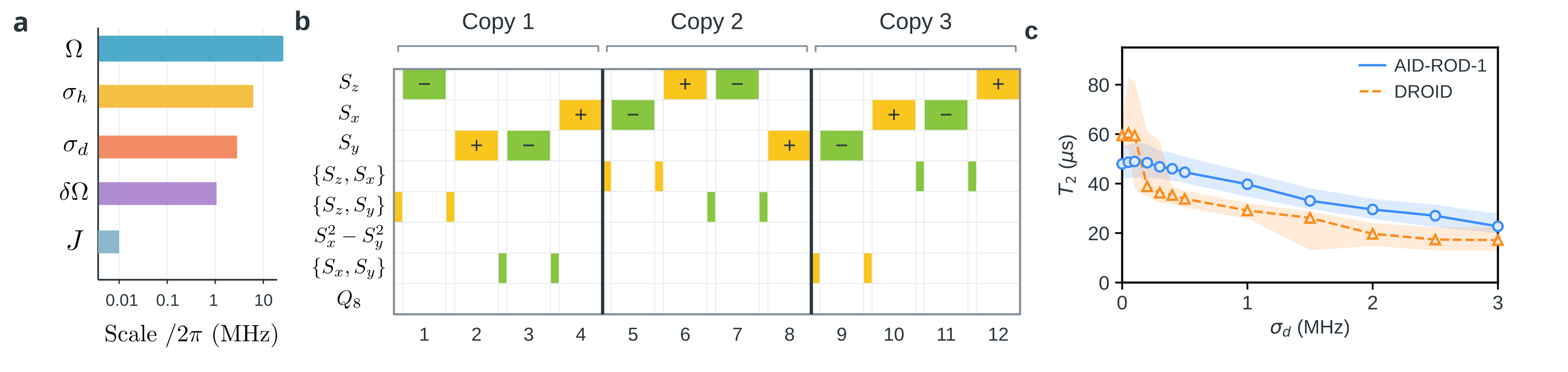}
    \caption{
    \textbf{(a)} Schematic frequency scales arranged vertically: per-tone drive amplitude $\Omega$; linear and quadratic detuning widths $\sigma_h,~\sigma_d$; drive-amplitude error during pulses $\delta\Omega$; and dipolar interactions $J$.
     Bar lengths indicate representative orders of magnitude.
    \textbf{(b)} Signed operator coefficients of $U_c^\dagger S_zU_c$ for the first 18-pulse block of AID-ROD-1. Labels Copy 1--3 denote the three consecutive repetitions of the same six-pulse unit within this block.
    Gold, green and white indicate positive, negative and zero coefficients, respectively.
    The spin dipole and quadrupole operators expand the spin-1 operator space and $Q_8 = \operatorname{diag}(1, -2, 1)/\sqrt3$.
    The toggling-frame trajectories of the three copies are related by conjugation with the ideal propagator $Q$ of the six-pulse unit, which satisfies $Q^3=I$. 
    \textbf{(c)} Performance of AID-ROD and DROID-C3PO as a function of the quadratic disorder strength $\sigma_d$, with the linear disorder width $\sigma_h = 4$ MHz and total NV density as $n \simeq 1.2$ ppm held fixed.
    }
    \label{<fig3>}
\end{figure*}

To understand the origin of AID-ROD’s advantage, we first compare the on-site disorder channels in spin-1 systems with those in their spin-1/2 counterparts.
For spin-1/2, diagonal on-site disorder reduces to a single detuning term proportional to $S_i^z$, apart from an irrelevant identity contribution.
Spin-1 additionally supports an independent quadratic term, $D_i(S_i^z)^2$, whose residual effects limit coherence under our experimental conditions.
Our analysis identifies improved suppression of this additional channel within individual 18-pulse blocks as one source of AID-ROD-1’s advantage.

The use of these short blocks reflects the separation of energy scales in our NV ensemble: onsite disorder and drive-amplitude errors can reach the megahertz scale, whereas the characteristic dipolar coupling is of order $J\sim2\pi\times10\ \mathrm{kHz}$ (Fig.~\ref{<fig3>}a).
Each block jointly suppresses linear and quadratic on-site disorder and common-mode drive-amplitude errors.
The seven blocks are then combined over a complete cycle to reduce residual dipolar interactions, assigning faster compensation to the dominant local errors.

To quantitatively interpret AID-ROD-1, we consider one complete cycle of seven 18-pulse blocks, $B_1,\ldots,B_7$.
Each block contains three identical copies of a six-pulse unit.
The six-pulse units in different blocks are related by cyclic shifts of their pulse--wait entries and fixed phase offsets on the two drive channels.
These transformations preserve the leading cancellation of the dominant local errors while reorienting residual interactions. 
We therefore analyze the first block, $B_1$, whose three copies are labeled Copy 1--3 in Fig.~\ref{<fig3>}b.

Let $Q$ denote the ideal propagator of one six-pulse unit in $B_1$.
In a fixed rephased spin basis, $Q^3=I$, with $O\mapsto Q^\dagger OQ$ cycling $S^z\to S^x\to S^y\to S^z$.
For \(H_{\rm loc}=hS^z+D(S^z)^2\), the three equally weighted copies within $B_1$ cancel the leading linear free-evolution contribution through $(S^x-S^z)+(S^y-S^x)+(S^z-S^y)=0$, and the corresponding quadratic terms yield only a global phase because $(S^x)^2+(S^y)^2+(S^z)^2=2\mathbb I$ (Fig.~\ref{<fig3>}b).
Including finite-pulse effects, the leading local residual is 
\begin{equation}
\overline H_{\rm loc}^{(0)}
=
-\frac{2\epsilon_{\Delta}}{3t_0}
\left(
\{S^z,S^x\}+\{S^z,S^y\}+\{S^x,S^y\}
\right),
\label{eq-AID-ROD-Average-Hamiltonian}
\end{equation}
where $t_0$ is the duration of one copy of the six-pulse unit, so the block duration is $3t_0$; $\epsilon_{\Delta} = \epsilon_+-\epsilon_-$,  and $\epsilon_\pm$ are the rotation-angle errors on the $\lvert0\rangle\leftrightarrow\lvert\pm1\rangle$ drive channels. Thus, only the differential channel error survives at this order; predominantly common amplitude errors, $\epsilon_+\simeq\epsilon_-$, are suppressed along with onsite disorder within one 18-pulse block.
Combining the seven blocks over a complete cycle further reduces the leading average dipolar interaction.
In contrast, DROID-C3PO does not fully cancel the finite-pulse contributions from quadratic disorder, $D_i(S_i^z)^2$, within its short blocks.
Their cancellation is deferred to symmetrization over a longer cycle~\cite{Zhou2024Qudit}.
Whereas this construction relies on geometric insights from decoupling frame graphs, QORA's agents discover locally robust blocks without requiring a predefined geometric picture.

With linear disorder and dipolar interaction strength fixed, spin-1 simulations show that AID-ROD-1 achieves longer coherence times than DROID-C3PO at larger quadratic-disorder strengths (Fig.~\ref{<fig3>}c).
We complement this spin-1 comparison with experiments in the effective spin-1/2 subspace of the NV ensemble spanned by $\lvert0\rangle$ and $\lvert-1\rangle$, where diagonal disorder reduces to a single linear term.
In this setting, the QORA-discovered sequence achieves coherence times comparable to those of the spin-1/2 DROID protocol.
Together with the quadratic-disorder simulations, this result supports the interpretation that improved suppression of the additional, independent quadratic disorder channel contributes to AID-ROD’s coherence advantage in the spin-1 manifold.

\section{Outlook}\label{sec12}

Our results demonstrate how a multi-agent AI system combining theoretical reasoning with automated experimentation can discover new quantum control protocols. Human interpretation connects their improved performance to the underlying error-suppression mechanisms, illustrating how collaboration between AI agents and human researchers can generate new physical insight.

Looking forward, our work opens the door to a few intriguing directions. 
First, a natural extension is to apply this approach to higher-dimensional quantum systems beyond spin-1, such as atomic, trapped-ion and superconducting qudits~\cite{Anderson2015SU16,Ringbauer2022SU7,Liu2023SU4,Tripathi2025QuditDD,Hrmo2023NativeQudit,Omanakuttan2023QuditEntanglers}. 
A central question is how robust pulse structures should change as the number and relative strengths of independent error channels increase. 
Adapting the search to each platform’s Hamiltonian and available controls could identify pulse blocks and compositions suited to different disorder and interaction regimes. 
Such extensions could reveal whether sequences discovered for different quantum platforms share common error-cancellation structures, providing transferable design principles for robust multi-level quantum control.

Second, an exciting opportunity in AI for science is to incorporate physical interpretation into the discovery loop \cite{Arlt2026MetaDesign,Cornelio2023AIDescartes}.
In our current work, human researchers identify the mechanisms underlying the sequences discovered by QORA.
Incorporating these explanations into agent reasoning and training could enable future systems to formulate competing physical hypotheses, select measurements that distinguish them and revise their models in response to experimental evidence. 
Closing this loop would allow physical understanding to guide new experiments and experimental discoveries to refine that understanding --- a step towards autonomous AI scientists for quantum research.

\vspace{2mm}

\emph{Acknowledgements}: 
This work is mainly supported by the U.S. Department of Energy DE-SC0026580.
We acknowledge additional support from the Center for Quantum Leaps at Washington University in St. Louis, and the National Science Foundation (QIS 2514391, 2503230, 2607484, and NRT LinQ 2152221). 
X. Lai acknowledges support from the Edwin Jaynes Fellowship in the Department of Physics at Washington University in St. Louis.

\bibliography{references}

\end{document}